\documentclass[prl,twocolumn,preprintnumbers,amsmath,amssymb]{revtex4}
\usepackage{graphicx}
\usepackage{color}

\begin{document}

\title{A bacterial ratchet motor}

\author{
R. Di Leonardo$^1$, 
L. Angelani$^2$, 
G. Ruocco$^{1,3}$,
V. Iebba$^4$,
M.P. Conte$^4$,
S. Schippa$^4$,
F. De Angelis$^5$,
F. Mecarini$^5$,
E. Di Fabrizio$^5$
}

\affiliation{
$^1$CNR-INFM CRS-SOFT, c/o Universit\`a di Roma ``Sapienza'', I-00185, Roma, Italy \\
$^2$CNR-INFM CRS-SMC, c/o Universit\`a di Roma ``Sapienza'', I-00185, Roma, Italy \\
$^3$Dipartimento di Fisica, Universit\`a di Roma ``Sapienza'', I-00185, Roma, Italy\\
$^4$Dipartimento Scienze di Sanit\`a Pubblica, Universit\`a di Roma ``Sapienza'', I-00185, Roma, Italy\\
$^5$BIONEM laboratory, Universit\`a della Magna Graecia, I-88100, Catanzaro, Italy
}

\begin{abstract}

Self-propelling bacteria are a dream of nano-technology.  These unicellular
organisms are not just capable of living and reproducing, but they can swim
very efficiently, sense the environment and look for food, all packaged in a
body measuring a few microns.  Before such perfect machines could be
artificially assembled, researchers are beginning to explore new ways to
harness bacteria as propelling units for micro-devices.  Proposed strategies
require the careful task of aligning and binding bacterial cells on synthetic
surfaces in order to have them work cooperatively.  Here we show that
asymmetric micro-gears can spontaneously rotate when immersed in an active
bacterial bath. The propulsion mechanism is provided by the self assembly of
motile {\it Escherichia coli} cells along the saw-toothed boundaries of a
nano-fabricated rotor.  Our results highlight the technological implications of
active matter's ability to overcome the restrictions imposed by the second law
of thermodynamics on equilibrium passive fluids.

\end{abstract}

\maketitle

The possibility of harnessing bacterial power opens up new and
interesting ways to generate motion at the micron scale. Pioneering work in
this field has been carried out in the past demonstrating the possibility of
propelling microstructures by permanently attaching a layer of motile bacteria
("bacterial carpet") on the surfaces of various synthetic objects like latex
beads or PDMS (Polydimethylsiloxane) structures \cite{darnton}. Phototactic
control of such bacterial actuated structures has been recently demonstrated
\cite{weibel, steager} by exposing localized regions of the swarm to ultraviolet light.
It has also been shown that motile bacteria can be permanently attached on
patterned microbeads resulting in random walks having diffusion coefficients
larger than thermal ones \cite{behkam}. A common drawback of the previous
approaches is that the resulting motions are random and unpredictable,
reflecting the chaotic dynamics of propelling bacteria. A major issue still
remains to be addressed: can we rectify the chaotic motions of bacteria to
drive a micro-machine in a unidirectional and predictable motion? Any realistic
hope to harness bacterial power to perform a predetermined task relies on that
possibility.  A first bacterial driven micro-motor was achieved by forcing the
unidirectional gliding of bacteria along circular tracks by a careful chemical
patterning of micro-structures \cite{hiratsuka, hiratsuka2}.  A microrotor was
then docked on the microtrack and observed to rotate unidirectionally at about
2 rpm.  

Here we demonstrate that the underlying off-equilibrium nature of a bacterial
bath allows to rectify the chaotic motions of bacteria by geometry alone and to
drive an asymmetric object into unidirectional motion by simple immersion in an
active bacterial suspension. Once the micro-structures are fabricated with a
proper asymmetric shape no further chemical patterning or externally induced
taxis is needed to produce a directional and predictable motion. 

Suspensions of self propelling bacteria can be viewed as a class of strongly
non-equilibrium fluids, often referred to as active matter \cite{vicsek, wu,
ramaswamy, hern, epj}.  Though at a first look, bacterial motions resemble the
chaotic kinetic dynamics of molecules in a gas, the underlying dynamics is
intrinsically irreversible.  The constant action of the flagellar rotary motor
results in a propelling force pushing the cell body against the fluid drag
force. Such propelling forces act as a non-conservative external force field,
precluding the possibility of a physical treatment based on the tools of
equilibrium statistical mechanics.  When a passive physical system is placed in
contact with such an active bath, the dynamics arising from energy transfers
can give rise to completely new phenomena. Equilibrium thermal baths are known
to induce Brownian motion on suspended objects, as the result of continuous
energy fluctuation and dissipation events, during which an external object
extracts kinetic energy from the bath, and releases it back to the bath through
viscous damping \cite{kubo}.  At thermal equilibrium, fluctuations involving
the same amount of energy have the same probability, and no matter how
asymmetric is the shape of an object, any state characterized by a given speed
occurs with the same probability of its time reversal. Brownian motion,
therefore sets no time arrow and looks exactly the same under time reversal.
Here we show that, once spatial inversion symmetry is broken, an
off-equilibrium active bath, in the absence of time reversal constraints imposed
by equilibrium, can provide strongly biased fluctuations. In practical terms we
demonstrate that giving an appropriate asymmetric shape to an object, a
spontaneous directed motion can result when the object is immersed in an active
bacterial suspension.

{\it E.coli} cells are propelled by thin helical filaments driven by
bidirectional rotary motors embedded in the cell wall \cite{Ecoli, flagella1,
flagella2}. When all motors rotate anticlockwise the cell body is propelled in
an almost linear run.  Forward runs are intercalated by randomly distributed
tumble events, during which one or more motors start rotating clockwise
unbundling flagella and causing a random reorientation of the cell body.
Although bacteria can sense the chemical environment and tumble less frequently
when swimming along nutrient gradients \cite{chemo}, mechanical constrains seem
to have no observable influence on the behavior of flagellar motors. As a
consequence, if their motion is pinned by a wall, they will keep on and push on
the wall.  The presence of concave regions in an object boundary can therefore
induce local ordering of incoming bacteria and make them push cooperatively in
the same direction.  Here we demonstrate that a nano-fabricated object, with
appropriate asymmetric shape, can be actually pushed in unidirectional rotation
by self organization of bacteria on its boundary.  

{\it E. coli} (MG1655) strain was grown overnight at 33$^\circ$C
in Brain Heart Infusion Broth (BHI, Oxoid, Italy) containing calf brain
infusion solid 12.5 g/l, beef heart infusion solids 5.0 g/l, proteose peptone
10.0 g/l, glucose 2.0 g/l, sodium chloride 5.0 g/l, di-sodium phosphate 2.5
g/l. Alternatively, MG1655 strain was grown at 33$^\circ$C in tryptone broth
(TB, Difco) containing 1\% tryptone and 0.5\% NaCl. The saturated culture was
then diluted 1:100 (50 $\mu$l in 5 ml) into fresh medium (BHI or TB) and grown
at 33$^\circ$C until OD600 = 1.1 (corresponding to 1 10$^9$ CFU/ml)  was
reached, corresponding to a middle-log phase. Proper experiments of
growth-curve assessment were achieved. Bacterial cells were harvested from a
500 µl culture media by centrifugation at 2200 rpm for 10 minutes at room
temperature, the pellet was resuspended by gently mixing (avoiding pipetting or
vortexing in order to circumvent flagella breakdown) in 100 $\mu$l of
pre-warmed (33¡C) physiologic saline solution (0.85\% NaCl) or, alternatively,
physiologic saline solution containing 10 mM potassium phosphate and 0.1 mM
Na-EDTA (pH 7.0) \cite{adler}. This process was repeated three times to achieve
growth medium depletion and a suitable final bacteria concentration (5 10$^9$
CFU/ml). 

A pattern of 100x100 micro-gears was written by e-beam  lithography onto a
Glass-Chromium substrate, coated by electron-resist (PMMA). The exposure dose
is 320 mJ/cm$^2$.  After resist development and selective chromium wet etching,
a negative pattern of 10000 gears was obtained on the glass mask.  A silicon
substrate is coated by a bi-layer of 50 nm PMMA (as sacrificial layer) and 15
$\mu$m SU-8 negative resist. A pattern of target objects is written on the
sample by UV-lithography  using an exposure density power of 22 mW/cm$^2$.
After optical lithography the structures are still attached onto the PMMA
sacrificial layer, which can be removed by rinsing and sonicating the whole
sample in acetone. Micro-gears are finally sedimented by centrifugation and
resuspended in water.

\begin{figure}[ht]
\centerline{\includegraphics[width=.5\textwidth]{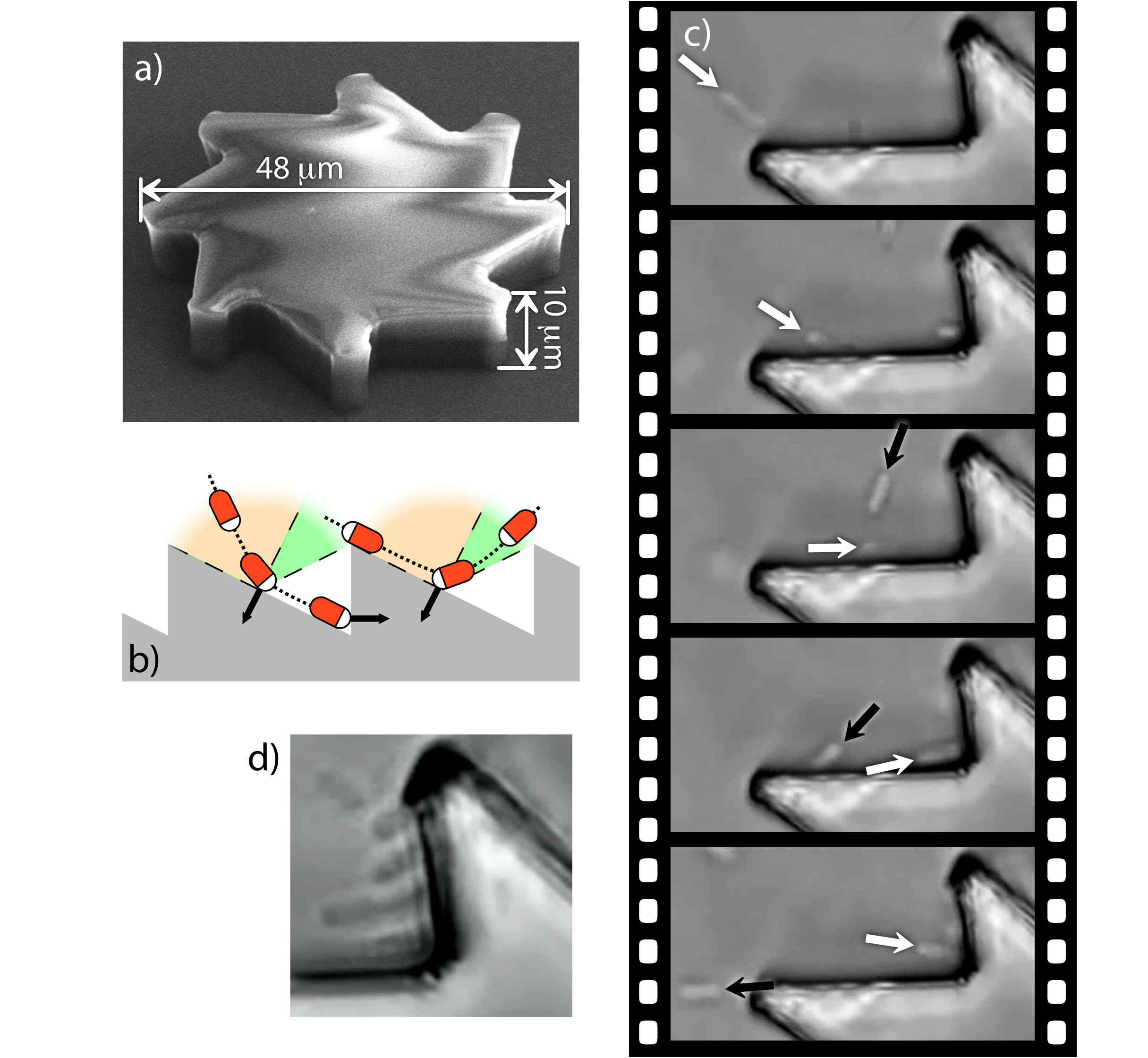}}
\caption{Rectifying bacterial motions with asymmetric boundaries.
a) An asymmetric polymeric gear is obtained by
optical lithography of a SU8 layer deposited over a PMMA sacrificial layer.
b) A pictorial representation of the mechanism through which the asymmetric shape
of the inclined teeth induces spontaneous alignment and locking
of self-propelled bacteria in the concave corners. Bacteria are drawn with a white head
pointing in the direction of self-propulsion. Black arrows represent the forces
exerted by bacteria on the walls. Equal and opposite reaction forces reorient the cell body
along the wall.
c) Two {\it E.coli} cells interacting with gear boundaries. The cell pointed by the white arrow
aligns parallel to the wall and slides towards the corner where it gets stuck, contributing a torque. 
The other one, pointed by the black arrow,
aligns along the wall and slides back into the bulk solution.  
d) Four bacterial cells spontaneously align and cooperatively push in the same direction against the wall.
}
\label{bound}
\end{figure} 

A first simple design is
that of a rotating micro saw-toothed disks having an external diameter of 48
$\mu$m and 10 $\mu$m thickness (Fig.  \ref{bound}a).  Low aspect ratio, quasi
planar shapes, are easily fabricated from SU-8 using optical lithography.  The
gears are then dispersed in a bacterial suspension of {\it E.coli} cells in a
motility buffer.  A drop of the resulting suspension hangs from the concave
part of a glass slide leaving an air gap between the liquid suspension and the
bottom coverslip (see supplementary material).  Gears then sediment on that
liquid-air interface where they are free from strong adhesion to solid
surfaces. Working at a free interface also reduces hydrodynamic coupling of
bacteria to the boundary surface. Interactions between cells and object
boundaries are well described by a repulsive contact force directed along the
surface normal. As a consequence, when a swimming cell hits a wall, it aligns
and slides parallel to it in a direction determined by its incoming angle
(Fig.\ref{bound} b,c and supporting movie). Cell sliding towards a concave
corner get stuck and start pushing until a tumbling event occurs reorienting
the cell and eventually setting it free to swim back into the solution.
Bacteria that got stuck at a corner are observed to remain in a "locked"
pushing configuration for a time interval varying from few seconds up to a
couple of minutes. In a concentrated suspension of bacteria, wall interactions
and inter-cellular repulsion lead to local packing and ordering of bacteria at
concave corners (Fig. \ref{bound}d). From visual inspection of recorded movies
we can estimate that an average of approximately 2 bacteria, among about 10
cells covering a single tooth, is contributing to the net torque. The
cooperative action of rectified bacteria results in a total torque which is
large enough to rotate our micro-gear, having a volume of approximately 20000
single cells, at an average angular speed of about 1 rpm.  Lying on a
liquid-air interface, half of the bulk drag force acting from the bottom
surface is now negligible so that the torque exerted on the gear can be
estimated assuming a rotational drag coefficient which is twice that of a disk
suspended in bulk buffer solution:

\begin{equation}
T \simeq \frac{16 \mu R^3}{3}\Omega \sim 10 \textrm{ pN $\mu$m} 
\end{equation}

where $\mu$ is the buffer viscosity, $R=20$ $\mu$m  is an average gear radius.

\begin{figure}
\includegraphics[width=.5\textwidth]{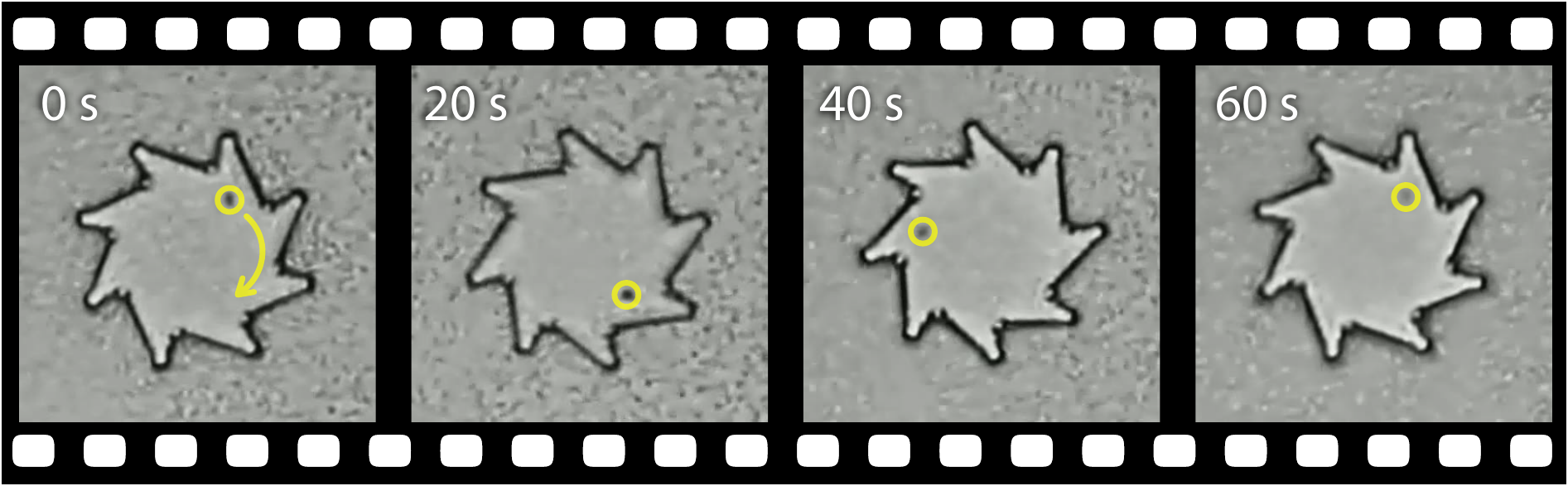}
\caption{A nano-fabricated asymmetric gear
(48 $\mu$m external diameter, 10 $\mu$m thickness) rotates clockwise at 1 rpm
when immersed in an active bath of motile {\it E.coli} cells, visible in the
background. The yellow circle points to a black spot on the gear which can be
used for visual angle tracking}
\label{spin}
\end{figure}

\begin{figure}
\includegraphics[width=.45\textwidth]{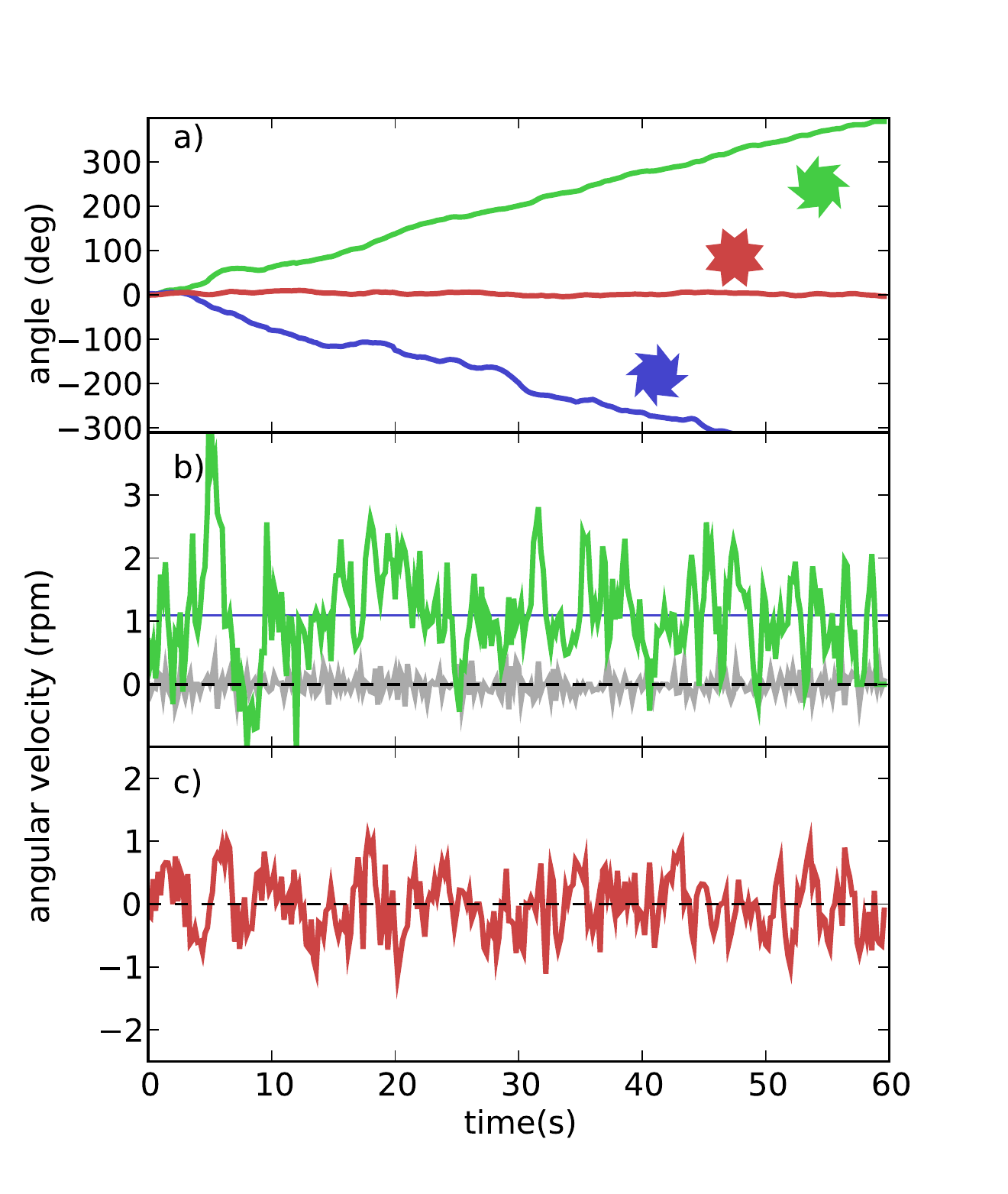}
\caption{a) Asymmetric gears rotate in a
bacterial bath while symmetric don't.  Green line is the time plot of the rotation angle of an
asymmetric gear moving in a quasi-steady rotation at 1.1 rpm. Blue line shows the opposite rotation
of a gear sedimented with inverted symmetry. Although the two datasets have been obtained with
bacteria from different cultures (BHI, TB), swimming in different buffers (unbuttered saline or motility buffer \cite{adler} ), 
the absolute value of rotation rates are comparable.
On the contrary, no net rotation is observed for the symmetric gear as shown by the red line.
b) The angular velocity of an asymmetric gear is reported as the green curve.
The blue solid line in the background represents the average angular speed of 1.1 rpm.
Fluctuations around the non-zero mean value are due to randomness in bacterial
number and local arrangement along the boundaries. The same gear in a bacteria free
buffer fluctuates much less as shown by the gray curve. The noise in the gray
data is mainly due to numerical errors in the algorithm for angular tracking.
c) A symmetric gear in a bacterial bath fluctuates with an angular velocity of
zero mean and a standard deviation comparable to the asymmetric case.}
\label{speed}
\end{figure}

We have filmed our rotary motor for one minute (Fig.2 and supporting movie) and
extracted angular velocity and rotation angle as a function of time (Fig.
\ref{speed}). Fig \ref{speed}a shows a comparison between the angular motions
of  asymmetric (green) and symmetric (red) gears in a bacterial bath.
Symmetric gears fluctuate with an average rotation angle that vanishes for
symmetry reasons.  Asymmetric gears break the spatial inversion symmetry of the
problem, and spontaneously rotate in a quasi steady motion. Fluctuations are
better appreciated in angular velocity plots Fig.  \ref{speed}b,c.  Asymmetric
and symmetric gears display quite large and comparable fluctuations in angular
velocities due to fluctuations in the number and local arrangement of bacteria
on the boundaries.  An asymmetric gear in a bacteria free buffer fluctuates
much less as shown by the gray line in Fig.  \ref{speed}b. Most of the noise in
the gray line comes from numerical errors in the angular video tracking
procedures.  Both mean velocities and fluctuations resemble very closely the
predictions of our molecular dynamics simulations \cite{angelani}. The swimming
velocity of a single bacterium provides the maximum achievable linear rotation
of the micro-gear edge. At higher rotation rates bacteria wouldn't be able to
catch up the moving walls.  In our case the linear speed of the edge is only
2.5 $\mu$m/s while bacteria swim at approximately 20 $\mu$m/s. This observation
leaves a large margin of improvement on the achievable rotation rates.
Asymmetric gears can sediment in two different orientations which are observed
to spin in opposite directions.  All these observations exclude the possibility
that rotation could be due to circular swimming of bacteria close to interfaces
\cite{diluzio}.  Bacterial adhesion to microgears seems to be negligible as can
be observed in the supporting movie illustrating the rectification action of
the sawtoothed boundary of the microstructures. The original movie lasts for
much longer and pushing bacteria are observed to swim away from the boundary
even after several minutes of contact. After a few hours, however, rigid
assemblies of bacteria start to form around the gear probably due to biofilm
formation.

The idea that, in non-equilibrium states, a directional motion can arise from
the chaotic dynamics of small molecules, was first put forward by Feynman in
his famous ``ratchet and pawl'' thought experiment \cite{flop}.  The
combination of asymmetry and non-equilibrium was soon recognized to be at the
origin of the ``ratchet effect'' opening the way to the stimulating concept of
Brownian motors in physical and biological contexts \cite{astumian}. Many ways
have been considered to drive a system out of thermal equilibrium, such as
cycling temperature or applying time-dependent external fields \cite{reimann}.
Our experiment demonstrates a novel realization of a ratchet mechanism, where
bacteria can be thought as intrinsically off-equilibrium ``molecules'' and an
asymmetric boundary is enough to generate directional motion.

There is still a lot to do to improve the performance of bacterial motors.  In
particular, different shapes and sizes are to be investigated as well as
different bacterial strains having different aspect ratios and different
swimming strategies and efficiencies.  However we can already foresee a
completely new technology, where passive micro-devices can be fabricated and
simply actuated by immersion in an active fluid. Applications at the micrometer
scale, such as self-propelling micro-machines or pumps and mixers for
micro-fluidics, are the most promising, but it will also be important to answer
the question whether bacterial motors are confined to the micro-world, or we
can think of a macroscopic exploitation of bacteria as mechanical power
sources.


\end{document}